\documentclass[apj]{emulateapj}
\lefthead{TSUJIMOTO et al.}
\righthead{MULTIPLE R-PROCESS EVENTS IN THE DRACO}

\def\ltsima{$\; \buildrel < \over \sim\;$}
\def\ltsim{\lower.5ex\hbox{\ltsima}}
\def\gtsima{$\; \buildrel > \over\sim \;$}
\def\gtsim{\lower.5ex\hbox{\gtsima}}
\def\ms{$M_{\odot}$ }
\def\msp{$M_{\odot}$}

\slugcomment{Accepted for publication in ApJ Letters}

\begin{document}
\title{Enrichment in r-process elements from multiple distinct events in the early Draco dwarf spheroidal galaxy\footnote{\scriptsize{This work is based on data collected at the Subaru Telescope, which is operated by the National Astronomical Observatory of Japan.}}}

\author{Takuji Tsujimoto\altaffilmark{1,2}, Tadafumi Matsuno\altaffilmark{2}, Wako Aoki\altaffilmark{1,2},  Miho N. Ishigaki\altaffilmark{3}, and Toshikazu Shigeyama\altaffilmark{4}
}

\affil{$^1$National Astronomical Observatory of Japan, Mitaka, Tokyo 181-8588, Japan; taku.tsujimoto@nao.ac.jp \\
$^2$Department of Astronomical Science, School of Physical Sciences, SOKENDAI (The Graduate University for Advanced Studies), Mitaka, Tokyo 181-8588, Japan \\
$^3$Kavli Institute for the Physics and Mathematics of the Universe (WPI), University of Tokyo, Kashiwa, Chiba 277-8583, Japan 
\\
$^4$Research Center for the Early Universe, Graduate School of Science, University of Tokyo, 7-3-1 Hongo, Bunkyo-ku, Tokyo 113-0033, Japan
}

\begin{abstract}
The stellar record of elemental abundances in satellite galaxies is important to identify the origin of r-process because such a small stellar system could have hosted a single r-process event, which would distinguish member stars that are formed before and after the event through the evidence of a considerable difference in the abundances of r-process elements, as found in the ultra-faint dwarf galaxy Reticulum II (Ret II). However, the limited mass of these systems prevents us from collecting information from a sufficient number of stars in individual satellites. Hence, it remains unclear whether the discovery of a remarkable r-process enrichment event in Ret II explains the nature of r-process abundances or is an exception. We perform high-resolution spectroscopic measurements of r-process abundances for twelve metal-poor stars in the Draco dwarf galaxy in the metallicity range of $-2.5<$[Fe/H]$<-2$. We found that these stars are separated into two groups with r-process abundances differing by one order of magnitude. A group of stars with high abundances of r-process elements was formed by a single r-process event that corresponds to the event evidenced in Ret II. On the other hand, the low r-process abundance group was formed by another sporadic enrichment channel producing a far fewer r-process elements, which is clearly identified for the first time. Accordingly, we identified two populations of stars with different r-process abundances, which are built by two r-process events that enriched gases at levels that differ by more than one order of magnitude.
\end{abstract}


\section{Introduction}

The recent progress in the research on r-process elements has led us to a better understanding of the production site of their elements. The detection of a near-infrared light bump in the afterglow of a short-duration $\gamma$-ray burst \citep[e.g.,][]{Tanvir_13} together with successful r-process nucleosynthesis calculations \citep[e.g.,][]{Wanajo_14, Rosswog_14, Goriely_15} has led researchers to focus on neutron star (NS) mergers as a promising site for r-process elements. The connection to a NS merger has been further strengthened from the perspective of Galactic chemical evolution \citep[e.g.,][]{Matteucci_14, Tsujimoto_14, Wehmeyer_15, Komiya_16}. Then, at last we have witnessed the compelling signature of r-process synthesis in the electromagnetic emissions associated with the gravitational-wave event, GW170817 which is an outcome of the binary NS meregr \citep[e.g.,][]{Smartt_17}. However, it still remains unsolved whether a NS merger is a unique channel of r-process production, as implied by an early galaxy enriched by r-process. A close look at the elemental feature of metal-poor stars provides no support for the delayed r-process enrichment, which is considered to be an inevitable outcome of NS merger events with the average merger time of $\sim$1 Gyr \citep{Dominik_12}; any metal-poor stars enriched by lighter elements such as $\alpha$-elements on a much shorter time-scale seem to be enriched by r-process as well \citep{Roederer_13}. 

The records of r-process elements deduced from the spectral lines of long-lived stars are a powerful tool for identifying the origin of their elements. One might think that such records can be searched at the Milky Way stars. However, according to the widely accepted theory of hierarchical galaxy formation scenario, the Milky Way was formed through the accretion of protogalactic fragments. This view predicts that the metal-poor stars in the Milky Way are an assembly of stars originated from individual fragments in which various chemical evolutions proceeded. Consequently, the observed abundance patterns should show a variety of patterns. This mechanism naturally explains the large variance observed in the abundances of neutron-capture elements, such as Eu, that are formed through the r-process at low metallicity \citep{Sneden_08}. Eventually, this explanation hampers our understanding of the origin and evolution of r-process elements. 

\begin{deluxetable*}{lrrcccrccr}
 \tablecaption{Summary of observations\label{tableobs}}
 \tablehead{\colhead{Object name} & \colhead{RA} & \colhead{DEC} & \colhead{Exposure} &\colhead{$S/N$\tablenotemark{a}} & \colhead{$V$\tablenotemark{b}} & \colhead{$K_s$\tablenotemark{c}} & \colhead{$E(B-V)$\tablenotemark{d}} & \colhead{$v_{\rm hel}$}\\
 & \colhead{(hh mm ss)} & \colhead{(dd mm ss)} & \colhead{(s)} & &  &  &  & \colhead{(km\,s$^{-1}$)}
}
  \startdata
Irwin 18126            & 17 20 41.82 & +58 00 24.8& 3600 & 37 & 17.07 & 13.77 & 0.031 & -304.0\\
Irwin 18525            & 17 20 37.39 & +57 59 12.6& 5400 & 36 & 17.15 & 14.03 & 0.030 & -278.6\\
Irwin 18645\tablenotemark{e} & 17 19 10.81 & +57 59 17.4& 4200 & 41 & 17.25 & 14.07 & 0.028 & -297.1\\
Irwin 18723            & 17 18 42.24 & +57 59 10.0& 5400 & 44 & 17.21 & 14.26 & 0.025 & -288.8\\
Irwin 19318\tablenotemark{e} & 17 21 03.55 & +57 57 08.2& 5400 & 44 & 17.39 & 14.28 & 0.029 & -288.9\\
Irwin 22391            & 17 20 43.67 & +57 48 44.2& 3300 & 54 & 16.87 & 13.33 & 0.024 & -285.7\\
Irwin 22509            & 17 19 47.67 & +57 48 36.6& 5400 & 48 & 17.28 & 14.15 & 0.027 & -280.6\\
Irwin 22697\tablenotemark{e}            & 17 21 40.36 & +57 47 32.0& 5400 & 34 & 17.35 & 14.33 & 0.027 & -291.5\\
SDSS J171627+574808     & 17 16 27.79 & +57 48 08.6& 5400 & 42 & 17.08 & 14.07 & 0.033 & -285.8\\
2MASS J17200577+5756234 & 17 20 05.78 & +57 56 23.4& 3600 & 32 & 17.15 & 14.37 & 0.028 & -290.7\\
\enddata
\tablenotetext{a}{Signal-to-noise ratio per $1.2\,\mathrm{km\,s^{-1}}$ pixel around $5852\,\mathrm{\AA}$}
\tablenotetext{b}{Converted from $g,\,r,\,i$ CFHT magnitudes of \citet{Segall2007}}
\tablenotetext{c}{2MASS magnitude from \citet{Cutri2003}}
\tablenotetext{d}{\citet{Schlafly2011}}
\tablenotetext{e}{Observed with $R\sim 60,000$}
\end{deluxetable*}

Alternatively, the small stellar masses in the satellite galaxies of the Milky Way present an advantage in the study of r-process. Relatively small number of stars that were formed in these galaxies allows the detection of rare potential r-process production events such as NS mergers \citep{Tsujimoto_14}. Recent reports on the ultra-faint dwarf (UFD) galaxy Reticulum II (Ret II) revealed that a single event with a high r-process yield considerably enhanced the r-process abundances of this galaxy \citep{Ji_16, Roederer_16}. However, these studies did not clarify whether the discovered event is a ubiquitous one or not. This uncertainty remains after one star in another UFD galaxy, Tucana III, was found to be enriched to a level similar to that of stars in Ret II \citep{Hansen_17}. We have no clues for identifying the r-process origin in Tucana III as the available information on r-process abundance is restricted to a single star.

The Draco dwarf spheroidal galaxy (dSph) is not as faint as Ret II, but is still a small stellar system that could be significantly affected by a single event to produce r-process elements in its early chemical evolution. \citet{Tsujimoto_15} found that relatively metal-rich ([Fe/H]$>-2$) stars in Draco have a constant abundance of Eu (i.e., [Eu/H] $\sim-1.5$), whereas more metal-poor stars have no detectable abundances of Eu. The low observed Ba/Eu ratios for stars with [Fe/H] $\leq -2$, including the metal-poor end of the constant Eu abundance, a feature which the Draco dSph has in common with the other dSphs such as Carina and Sculptor \citep{Tsujimoto_15} guarantee that Ba and Eu in the observed stars were produced by the r-process \citep{Simmerer_04, Bisterzo_14}. Thus, this Eu abundance feature suggests that the Draco dSph experienced a large increase in r-process abundance within a small [Fe/H] range. However, the data size and quality are still too limited to exclude the possibility of a gradual increase in r-process elements with increasing amounts of Fe. 

Here we report the result of Y, Ba, and Eu abundances from analysis of high-resolution spectra for 12 stars, 10 stars of which are newly measured, in the narrow metallicity range of $-2.5<$[Fe/H]$<-2$. Thanks to our strategy to target the stars around the metallicity where r-process event is anticipated to occur, we can assess how the abundance of r-process elements changed at this early stage of Draco dSph's formation.

\section{Observation}

We performed observations with the High Dispersion Spectrograph \citep[HDS; ][]{Noguchi2002} on the Subaru Telescope during May 29 - June 01, 2016. Details of the observation are summarized in Table 1. We selected 10 red giants in Draco in the metallicity range of $-2.5<$[Fe/H]$<-2$, as estimated by the spectroscopy at medium-resolution \citep{Kirby2010} and moderately high resolution \citep{Walker2015}. We adopted a standard setup for HDS, which covers $4000-6800\,\mathrm{\AA}$, and applied $2\times 2$ CCD binning. Most of our targets were observed with a spectral resolution of $R\sim 45,000$, and three stars were observed with $R\sim 60,000$. In addition to 10 stars, we reanalyzed two stars from \citet{Tsujimoto_15} with [Fe/H] values in the abovementioned range. Data reduction was performed with an HDS reduction pipeline that utilizes IRAF\footnote{IRAF is distributed by the National Optical Astronomy Observatory, which is operated by the Association of Universities for Research in Astronomy, Inc. under a cooperative agreement with the National Science Foundation.} 

\section{Abundance analysis}

\begin{figure*}[t]
\vspace{0.4cm}
\begin{center}
\includegraphics[angle=90,width=14cm,clip=true]{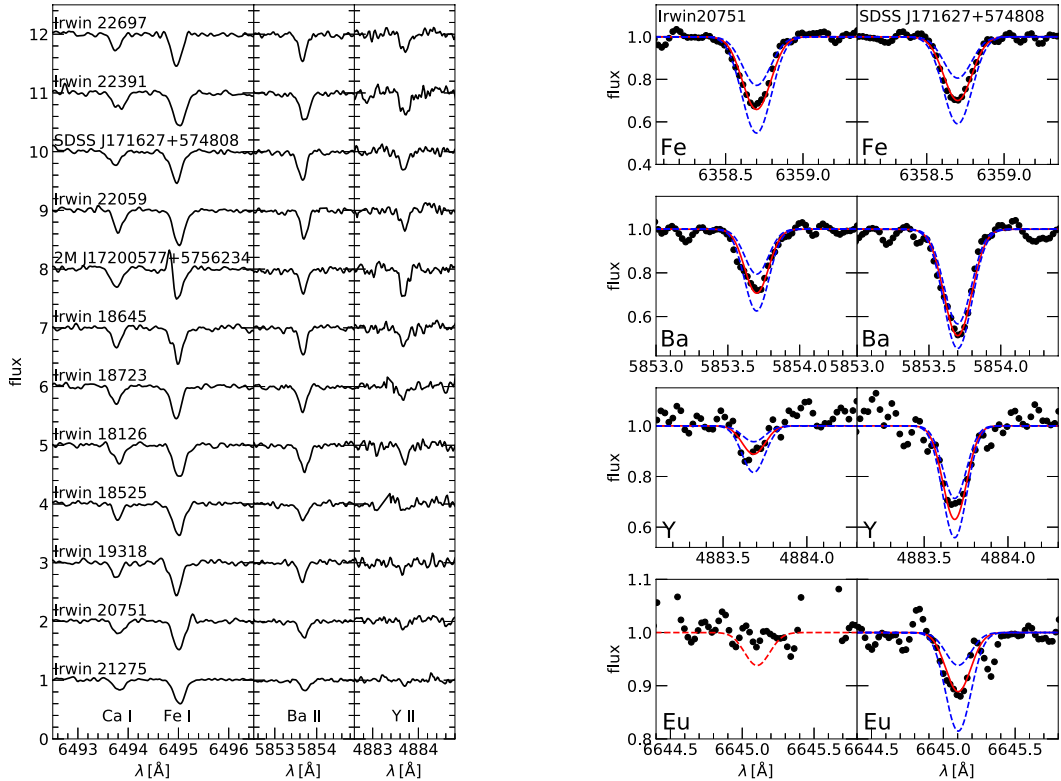}
\end{center}
\caption{{\it left panel}: The spectral region around the Ca, Fe, Ba, and Y absorption lines for the 12 stars in Draco. Although the Ca and Fe lines are quite similar, the strengths of the Ba and Y lines differ markedly for the eight objects at the top vs. the other four objects. {\it right panel}: The absorption lines of Fe I $6358\,\mathrm{\AA}$, Ba II $5854\,\mathrm{\AA}$, Y II $4884\,\mathrm{\AA}$, and Eu II $6645\,\mathrm{\AA}$ of Irwin 20751 (left panels) and SDSS J171627+574808 (right panels). The two stars have similar stellar parameters but show the different Ba, Y, and Eu abundances. The observed spectra indicated by black points are compared with the best fit results (red solid lines), together with the synthesized spectra with $\Delta$[X/H]=$\pm$0.3 dex offsets (X=Fe, Ba, Y, Eu; blue dashed lines). The red dashed line for Eu in Irwin 20751 shows a synthetic spectrum to determine the upper limit of Eu abundance.
}
\end{figure*}

The effective temperatures were derived from $V-K_s$ color \citep{Segall2007, Cutri2003} using calibration of \citet{Casagrande2010} with correction for interstellar extinction \citep{Schlafly2011}. To determine surface gravities,  assuming that each star is on the red giant branch, we used Yonsei-Yale isochrones \citep{Kim2002} with an age of $10.9\,\mathrm{Gyr}$ \citep{Orban_08}. Although some stars needed an extrapolation of the isochrones toward a lower surface gravity, the abundance measurement using Fe II lines resulted in Fe abundances consistent with those deduced from Fe I lines within $\Delta [\mathrm{Fe/H}]=0.15\,\mathrm{dex}$. The micro-turbulent velocities $v_t$ were determined so that the Fe abundances from individual Fe lines show no correlations with the strengths of the lines. We adopted the uncertainties for stellar parameters as follows. Uncertainties for the effective temperatures originated from photometric errors were in the range of $60-100\,\mathrm{K}$. Here we adopted $100\,\mathrm{K}$ as $\sigma(T_{\rm eff})$ for all the stars. The value of $\sigma(\log g)$ was determined as $\sigma(\log g)=0.4\,\mathrm{dex}$ from the typical uncertainties of $T_{\rm eff}$ and metallicity. The errors on $v_t$, $\sigma(v_t)\sim 0.2\,\mathrm{km \ s^{-1}}$, were estimated from the range of its values that give no apparent trend between line strengths and derived abundances. We adopted $\sigma(v_t)= 0.3\,\mathrm{km \ s^{-1}}$ as the uncertainty, considering an additional error budget arised from degeneracy between $v_t$ and $T_{\rm eff}$.

We performed an abundance analysis under 1D/LTE approximation with the ATLAS NEWODF grid of alpha-enhanced model atmospheres \citep{Castelli_03} and the spectrum synthesis code used in \citet{Aoki_09}, which is based on the same assumptions as the model atmosphere program of \citet{Tsuji_78}. The Fe abundances were derived from equivalent widths of Fe I lines using the same line list as \citet{Tsujimoto_15}. The abundances of Ba and Eu were estimated by fitting the synthetic spectra to each absorption feature: Ba II lines around $5854\,\mathrm{\AA},\,6142\,\mathrm{\AA}$, and $6497\,\mathrm{\AA}$ and an Eu II line around $6645\,\mathrm{\AA}$. Hyperfine splitting was considered for Ba II and Eu II, by assuming the same isotope ratios of the r-process component that appear in the solar-system material, as done by \citet{McWilliam1998} and \citet{Ivans2006}, respectively. We determined Y abundances from Y II lines with atomic data from \citet{Hannaford1982}. Observed spectra for our 12 stars as well as comparisons between synthesized spectra and observed ones for two stars are shown in Figure 1. The uncertainties in the derived abundances were estimated with the standard deviations of the abundances from the individual lines divided by the square root of the number of lines. If the number of detected lines for the species was less than three, their standard deviations were replaced by those of the abundances from Fe I lines. In addition, we considered the effect of uncertainties of stellar parameters quadratically. Stellar parameters and derived abundances are summarized in Table 2. 

\begin{deluxetable*}{lrrcrrrrrr}
 \tablecaption{Stellar parameters and derived abundances\label{tableabundance}}
  \tablehead{\colhead{Object name}            &\colhead{$T_{\rm eff}$} & \colhead{$\log g$} & \colhead{$v_t$}   & \colhead{$[\mathrm{Fe/H}]$} & \colhead{$[\mathrm{Y/H}]$} & \colhead{$[\mathrm{Ba/H}]$} & \colhead{$[\mathrm{Eu/H}]$} & \colhead{$[\mathrm{Ba/Eu}]$}\\
              &\colhead{(K)} & \colhead{(dex)}  & \colhead{$\mathrm{(km\,s^{-1})}$}   & \colhead{(dex)} & \colhead{(dex)} & \colhead{(dex)} & \colhead{(dex)} &  \colhead{(dex)}
             }
  \startdata
  \multicolumn{9}{c}{high Ba stars}\\
Irwin 18126              & 4076 & 0.00 & 1.93 &$-2.37 \pm 0.18$& $-2.83\pm0.13 $  & $-2.15 \pm  0.23$&$-1.77    \pm  0.25$ & $-0.38 \pm 0.26$ \\
Irwin 18645              & 4223 & 0.54 & 1.90 &$-2.01 \pm 0.18$& $-2.66\pm0.13 $  & $-1.99 \pm  0.26$&$-1.55    \pm  0.26$ & $-0.44 \pm 0.32$ \\
Irwin 18723              & 4343 & 0.73 & 1.94 &$-2.09 \pm 0.18$& $-2.71\pm0.16 $  & $-2.05 \pm  0.26$&$-1.57    \pm  0.24$ & $-0.48 \pm 0.26$ \\
Irwin 22391              & 4096 & 0.02 & 2.21 &$-2.18 \pm 0.18$& $-2.36\pm0.17 $  & $-1.82 \pm  0.25$&$-1.29    \pm  0.27$ & $-0.53 \pm 0.31$\\
Irwin 22509              & 4232 & 0.55 & 1.85 &$-2.04 \pm 0.18$& $-2.38\pm0.11 $  & $-1.91 \pm  0.24$&$-1.42    \pm  0.26$ & $-0.49 \pm 0.29$\\
Irwin 22697              & 4255 & 0.47 & 1.46 &$-2.09 \pm 0.18$& $-2.61\pm0.14 $  & $-1.78 \pm  0.23$&$-1.66    \pm  0.28$ & $-0.12 \pm 0.30$\\
SDSS J171627+574808     & 4272 & 0.35 & 1.96 &$-2.34 \pm 0.18$& $-2.37\pm0.16 $  & $-1.87 \pm  0.25$&$-1.52    \pm  0.24$ & $-0.35 \pm 0.28$\\
2MASS J17200577+5756234 & 4510 & 1.09 & 1.89 &$-2.01 \pm 0.18$& $-1.90\pm0.14 $& $-1.98 \pm  0.26$&$-1.59    \pm  0.30 $ & $-0.39 \pm 0.35$\\
 \multicolumn{9}{c}{low Ba stars}\\
Irwin 18525              & 4238 & 0.40 & 1.98 &$-2.34 \pm 0.18$& $<-3.12$  & $-2.75 \pm  0.25$&$<-1.78            $ & --- \ \ \ \ \ \ \\
Irwin 19318              & 4232 & 0.29 & 1.89 &$-2.32 \pm 0.18$& $<-3.12$  & $-2.86 \pm  0.23$&$<-1.72            $ & --- \ \ \ \ \ \ \\
Irwin 20751              & 4260 & 0.36 & 2.08 &$-2.36 \pm 0.18$& $-3.35\pm0.21 $  & $-2.90 \pm  0.23$&$<-1.82            $ & --- \ \ \ \ \ \ \\
Irwin 21275              & 4413 & 0.67 & 2.10 &$-2.43 \pm 0.18$& $-3.61\pm0.22 $  & $-3.05 \pm  0.23$&$<-1.87            $ & --- \ \ \ \ \ \ \\
   \enddata
\end{deluxetable*}

We classify the stars into two groups in terms of [Ba/H]. For Ba-rich stars, we can deduce the [Ba/Eu] ratio, which shows no correlation with [Fe/H] and relatively low values with an average of $-0.45$,  indicating that Ba in our target stars originated mainly in r-process nucleosynthesis \citep{Simmerer_04, Bisterzo_14}. Our deduced ratio is slightly higher than the values found for r-process-enhanced stars, such as CS22892-052 \citep[=$-0.65$;][]{Sneden_03}. To check whether the slightly higher [Ba/Eu] ratios obtained for the high-Ba stars are caused by our abundance analysis, we carried out the analysis with the same procedure used in this work for a bright metal-poor r-process-rich star, HD221170. Stellar parameters (Teff, log g, [Fe/H]) and Eu abundances of this star are similar to those of our Draco target stars. We obtained [Ba/Eu]=$-0.51$, which is in good agreement with the value of $-0.54$ deduced from the detailed analysis by other study \citep{Ivans2006}. At the moment, the possibility of inclusion of Ba from s-process operation to some extent can not be excluded. Even in this case, the contribution from s-process has a minor effect on Ba abundance within $\sim$0.2 dex and we may conclude that a large part of Ba in our stars is representative of a r-process origin. 

\section{Two r-process populations}

Our results on the Ba abundance measurements are shown in the top panel of Figure 2. The stars were found to have separated into two groups in terms of their [Ba/H] ratios. Each mean [Ba/H] value is $\langle$[Ba/H]$\rangle$=$-2.89$ and $-1.92$, suggesting the occurrence of an episode boosting [Ba/H] by one order of magnitude around the  lowest metallicity in the high Ba abundance group (i.e., [Fe/H]$\approx-2.3$). These two distinct populations of the r-process-enriched stars are also implied from their Eu abundances. As shown in the top panel of Figure 3, Eu abundances were measured for all the stars belonging to the high-[Ba/H] ($>-2.2$) population, resulting in a mean value of $\langle$[Eu/H]$\rangle$=$-1.46$, while the spectral lines of Eu have not been detected ([Eu/H]\ltsim$-1.8$) for the low-[Ba/H] ($<-2.7$) population.   

Interestingly, the abundance of Y, a light neutron-capture element, follows a pattern similar to those of Ba and Eu (the bottom panel of Fig.~3). Two stars in the low-Ba population have [Y/H]$\approx-3.5$, and the other two stars show no detectable lines of Y. In contrast, the high-Ba population shares a similar value, i.e., [Y/H]$\approx-2.5$, suggesting a sudden increase in the [Y/H] ratio, which is as large as that seen for [Ba/H]. Therefore, it is likely that this r-process event is associated with a short period of increased production of both heavy and light neutron-capture elements. The average [Y/Ba] ratios for the high- and low-Ba populations are quite similar at approximately $-0.6$ and $-0.5$, respectively. These ratios broadly correspond to the lower bound of [Y/Ba] values among the Galactic metal-poor halo stars, which implies that all these elements are of pure r-process origin.

The large increase in r-process abundance found in the Draco dSph is reminiscent of a similar discovery in the UFD galaxy Ret II \citep{Ji_16}. The feature discovered in Ret II suggests that a single event with a high r-process yield remarkably enhanced the abundances of r-process elements in this galaxy. We found similarities between the two phenomena. First, a single event in both galaxies increased [Ba/H] up to approximately the same level, i.e., [Ba/H]$\sim -2$. In addition, this increase in the [Ba/H] ratio is accompanied by an increase in the abundance of light neutron-capture elements, such as Sr and Y,  to the same degree as the increase in Ba \citep{Roederer_16}. Such similarities in r-process abundance features suggest that the r-process event in Draco may be identical to that in Ret II. Accordingly the r-process event discovered in Ret II is not unique; however, it occurred in another Milky Way satellite, the Draco dSph. We note that the reported increase in the [Ba/H] ratio of as much as $\sim$2.5 dex in Ret II was caused by a considerably low [Ba/H] value of approximately $-4.5$ prior to this r-process event. Another difference between the two systems is the metallicity at the time of their occurrences, i.e., [Fe/H]$\approx -2.3$ in Draco and $-3$ in Ret II.

\begin{figure}[h]
\vspace{0.2cm}
\begin{center}
\includegraphics[width=8.cm,clip=true]{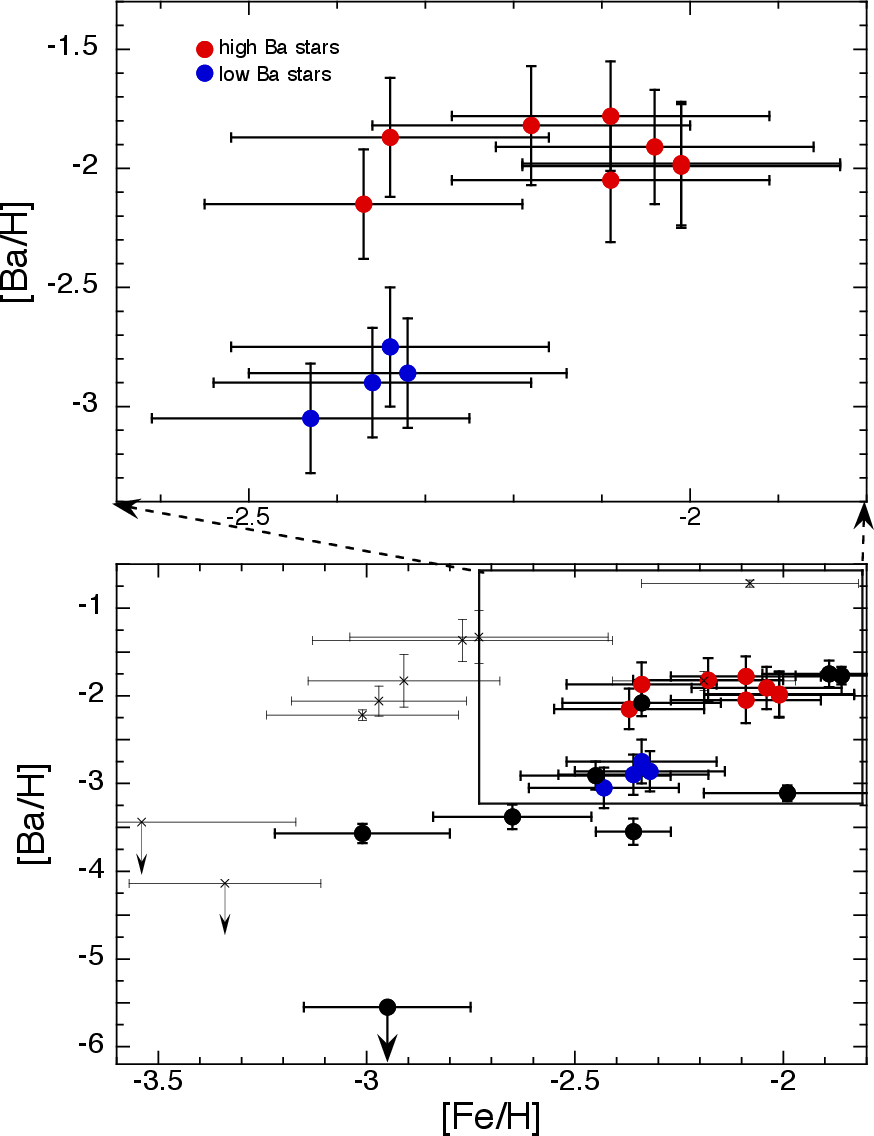}
\end{center}
\caption{{\it top panel}: Observed Ba abundances vs. Fe abundances for the 12 Draco stars. Our results are presented by dividing the stars into two groups: one with high Ba abundance ([Ba/H]$>-2.2$; red circles) and the other with low Ba abundance ([Ba/H]$<-2.7$; blue circles). {\it bottom panel}: Observed Ba abundance extending to a lower-metallicity region of Draco,  together with for the stars in Ret II \citep[crosses;][]{Ji_16}. We include the previously reported results \citep{Shetrone_01, Fulbright_04, Cohen_09} for the Draco stars that share their [Fe/H] ratio with ours and the three stars with [Fe/H]$<-2.5$. These data are indicated by black circles.
}
\end{figure}

We estimated the mass of Ba ejected from the r-process event found in the Draco dSph. By modeling the chemical evolution of Draco \citep{TsujimotoN_15}, we deduced that the mass fraction converted from the initial gas to stars was approximately15\%. Furthermore, from the current stellar mass of 3.2$\times 10^5$\ms \citep {Martin_08}, we obtained a mass of 2.1$\times 10^6$\ms as the initial mass of the gas in the Draco dSph progenitor, which would have been mixed with the ejecta that were enriched with r-process elements. Note that the gas mass used for star formation prior to the r-process event is negligible. Accordingly, to increase [Ba/H] from $-2.9$ to $-1.9$, we estimated that a single r-process event should have ejected 3$\times10^{-4}$\ms of Ba. This result can be compared with the theoretical yield from potential r-process producing events. An NS merger is predicted to typically eject 0.01 \ms of matter \citep[e.g.,][]{Barnes_13}. If we assume that the ejecta comprise elements with $A\geq$ 88, including Sr and Y, the expected mass of Ba is 3.2$\times 10^{-4}$\msp, which is in good agreement with our estimate from the Draco observation. We note that a black hole-NS merger can release a similar or greater mass of r-process elements \citep{Kyutoku_15}. Besides such mergers, another proposed candidate is peculiar core-collapse supernovae (CCSNe) that are characterized by fast rotations and high magnetic fields \citep[e.g.,][]{Takiwaki_09}. These magneto-rotational CCSNe (MR-SNe) can produce light and heavy r-process elements together with lighter elements such as $\alpha$-elements and Fe, and thus may contribute to their enrichment during early star formation in dSphs \citep{TsujimotoN_15}. The latest results on nucleosynthesis in MR-SNe suggest Ba production as much as $2.8\times10^{-4}$\ms in a model with low neutrino luminosity \citep{Nishimura_17}, which also agrees well with our estimate. The progenitor mass of Ret II \citep[2360 $L_\odot$:][]{Simon_15} could be comparable to that of Draco, i.e., approximately $10^6$\ms according to the assembling  mini-halo model \citep[e.g.,][]{Ji_15, Ji_16}. This mechanism may give the threshold of an initial mass of protogalaxies and would lead to a similar degree of r-process enrichment. 

\section{Two distinct r-process events}

In this section, we inspect the Ba feature for [Fe/H]\ltsim$-2.5$ from a handful of samples, including the results of previous studies \citep{Shetrone_01, Fulbright_04, Cohen_09}, to analyze how the Draco dSph was enriched up to [Ba/H]$\approx-2.9$ until the metallicity increased to [Fe/H]$\sim -2.5$. The bottom panel of Figure 2 demonstrates that [Ba/H] jumps from an undetectably low abundance \citep[$<-5.6$;][]{Fulbright_04} to approximately $-3.5$ around [Fe/H]=$-3$ \citep{Cohen_09}. Although the two stars of Draco showing [Ba/H]$\sim -3.5$ in the metallicity range of $-3<$[Fe/H]$<-2.5$ have a slightly different level of [Ba/H] from our low-Ba stars ([Ba/H]$\sim -2.9$), we consider them as the members of the low-Ba group, given the lack of statistics as well as possible observational errors. The existence of the low-Ba group suggests that the first r-process event occurred around [Fe/H]=$-3$ in Draco and induced a large boost in Ba abundance. Note that the r-process population prior to the first event is the lowest in terms of [Ba/H] ($<-5.6$), which is implied from the observation of the star Draco 119 and shares its record with the most iron-poor halo star in the Milky Way ([Ba/H]$<-6.1$; Keller et al.~2014). We emphasize that the first event, which lifted [Ba/H] to approximately $-3.5$, would have ejected a considerably smaller amount of r-process elements, as little as $<$10\% of the amount ejected by the event around [Fe/H]=$-2.3$. 

\begin{figure}[t]
\vspace{0.2cm}
\begin{center}
\includegraphics[width=8.cm,clip=true]{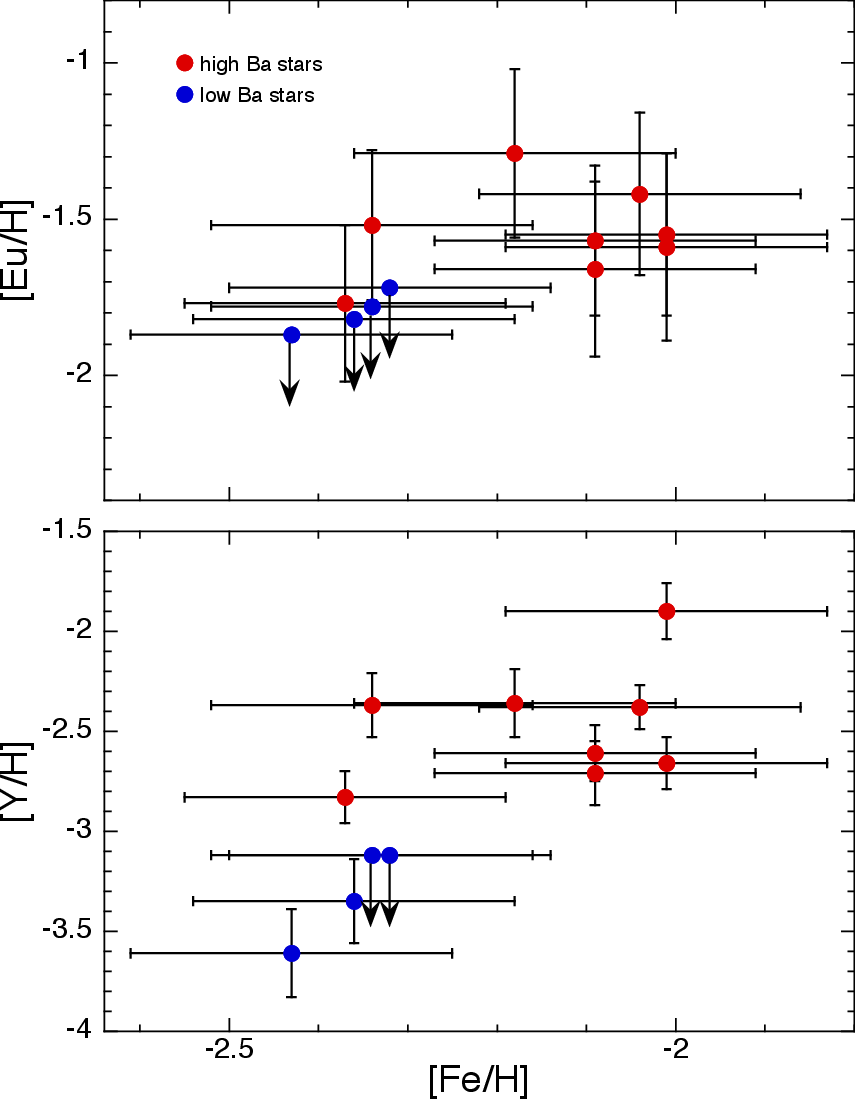}
\end{center}
\caption{{\it top panel}: The observed Eu ratios of our 12 Draco stars. Symbols are the same as those used in Fig.~2. {\it bottom panel}: Same as the top panel but for Y, a light neutron-capture element.
}
\end{figure}

As reported by previous studies \citep{Tsujimoto_14, Tsujimoto_15}, little r-process enrichment proceeds once [Fe/H]$>-$2, as implied by the constant Eu abundance ([Eu/H]$\sim-1.5$) demonstrated by classical dSphs including Draco (note that Ba abundance shows an increasing feature with an increase in Fe abundance due to s-process enrichment for this metallicity range). This finding does not exclude the possibility that an r-process event with an inefficient enrichment may have occurred, such as the first r-process event when [Fe/H]\gtsim$-2$. Nevertheless, such events are expected to be infrequent since UFDs including the relatively massive Hercules \citep[$3.7\times10^4 M_\odot$:][]{Martin_08}, do not generally show any features with Ba higher than the nearly constant value of  [Ba/H]$\sim-4.5$ \citep[e.g.,][]{Tsujimoto_14}. Accordingly, the pattern of r-process abundance can  be summarized as follows: the first r-process event occurs around [Fe/H] = |3, lifting the Ba abundance up to [Ba/H $\sim -3$ or $-3.5$. Subsequently one or more additional r-process events occur between $3<$[Fe/H]$< -2$, depending on the [Ba/H] ratio achieved by the first event. These events together lift the Ba abundance up to [Ba/H]$\sim -2$ and eventually build to the current r-process element abundance. 

We discovered two types of r-process events that differ by at least one order of magnitude in the degree of enrichment of interstellar matter. One of the two events, which produced a far fewer r-process elements, is not found in Ret II. The presence of an event building the low-Ba ([Ba/H]$\sim-2.9$) population that we discovered suggests that r-process events could provide the varying amount of nucleosynthesis products. This variation may suggest two distinct r-process sites, i.e., a NS merger and a MR-SN, in the early Draco. Alternatively, NS mergers would result in a wide range of amount of r-process elements owing to a large variety of ejecta mass, which is suggested by the diversity of brightness of radioactive kilonova \citep{Gompertz_17}. The NS merger rate estimated by the GW170817 event \citep{Abbott_17} might be compatible with this view since its rate deduced from the observed r-process abundance feature is smaller by approximately ten times, given all NS mergers yield high r-process production \citep{Tsujimoto_14}. Unveiling the origin of the two populations with high- and low-Ba abundances in the Draco dSph should be a priority in understanding the origin of r-process elements. Future research could approach this question with similar measurements of r-process elements in other classical dSphs. 

\acknowledgements

We are grateful to K. Aoki and staff members of the Subaru telescope for their helpful support and assistance in our HDS observation. This work was supported by JSPS KAKENHI Grant Numbers 15K05033, 15H02082, 16H02168, and 16H06341.

\end{document}